\documentclass[aps,prl,twocolumn,groupedaddress]{revtex4-2} 

\usepackage{pgfplots}
\usepackage{pgfplotstable}
\usepackage{filecontents}
\usepackage{placeins}
\usepackage{footnote}
\usepackage[para]{footmisc}
\usepackage{float}

\begin{document}


\title{Particle Builder A Board Game for the Teaching of the Standard Model of Particle Physics at a Secondary Level.}




\author{Lachlan McGinness}
\affiliation{Australian National University and Commonwealth Scientific and Industrial Research Organisation}

\author{Yutong Ma}
\affiliation{Australian National University}

\author{Mohammad Attar}
\affiliation{Australian National University}

\author{Andrew Carse}
\affiliation{Australian National University}

\author{Yeming Chen}
\affiliation{Australian National University}

\author{Thomas Green}
\affiliation{Australian National University}

\author{Jeong-Yeon Ha}
\affiliation{Australian National University}

\author{Yanbai Jin}
\affiliation{Australian National University}

\author{Amy McWilliams}
\affiliation{Australian National University}

\author{Theirry Panggabean}
\affiliation{Australian National University}

\author{Zhengyu Peng}
\affiliation{Australian National University}

\author{Jing Ru}
\affiliation{Australian National University}

\author{Jiacheng She}
\affiliation{Australian National University}

\author{Lujin Sun}
\affiliation{Australian National University}

\author{Jialin Wang}
\affiliation{Australian National University}

\author{Zilun Wei}
\affiliation{Australian National University}

\author{Jiayuan Zhu}
\affiliation{Australian National University}





\maketitle

\section{Introduction}
\vspace{-0.1cm}
Over the last decade, the Standard Model of Particle Physics has been included in many secondary physics curricula worldwide \cite{KranjcHorvat2022What}. However, its abstract nature and heavy reliance on symbolic representation make it difficult to teach at the secondary-school level. In response, a number of hands on resources have been developed to teach specialised lessons on scattering \cite{CERN2020Scattering}, anti-matter \cite{McGinness2019Printable} and quarks \cite{McGinness2019ThreeD}. The Particle Builder board game was created in 2016, by an international group of physics teachers. The game introduces the structure of matter and core ideas of particle physics through interactive gameplay, allowing students to build particle systems, compare their properties, and discover the underlying rules of the Standard Model themselves.\\

The physical version of the board game is freely available on CERN's Zenodo repository (https://zenodo.org/records/3594204) and has been viewed over 11,000 times and downloaded over 2000 times \cite{McGinness2018Particle}. 
However, there is a significant time cost for classroom teachers to print sufficient copies of the game for a class to play. To address this problem, the game was later adapted into a browser-based format that allows students to play against a basic Artificial Intelligence (AI) \cite{McGinness2024Particle,Attar2025Particle}, available at https://particle-builder.anu.edu.au. \\

We have tested the game with 281 Australian high school students. Of these, 225 completed both a pre-test and post-survey, allowing us to evaluate their learning outcomes and collect feedback on the game. We use these data to determine whether the game can improve student understanding of the Standard Model.

\section{The Particle Builder Game}
\vspace{-0.1cm}
There are seven `game' levels in Particle Builder with different rules. Level one is the simplest, while level seven contains the most analogies to particle physics concepts. In the most basic version of the game, students collect cards, which represent fundamental particles, to build a particle system (`target'). \\

\begin{figure}[!htbp]
	\includegraphics[width=0.4\textwidth]{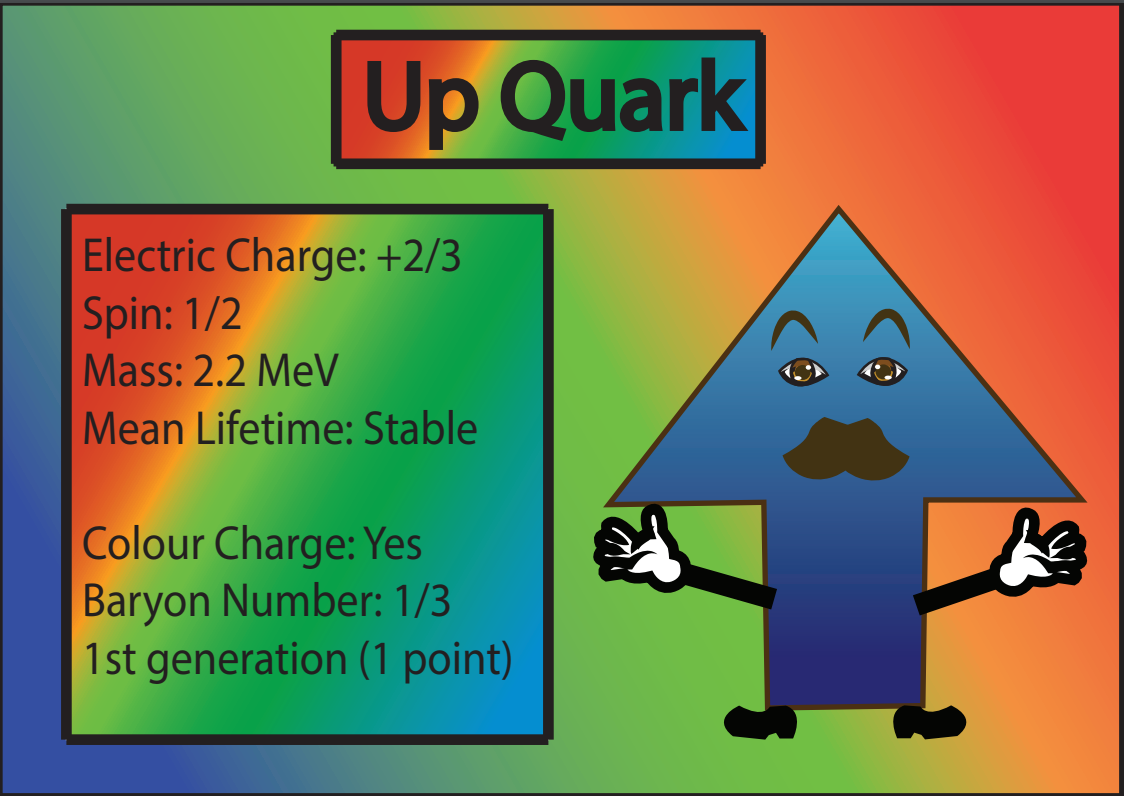}%
	\caption{\label{fig:UpQuark} The Up Quark card from Particle Builder. The card contains the electric charge, spin, mass and mean lifetime which the students must examine regularly in the game. Electric charge is given in units of elementary charge (e).}
\end{figure}

Each card lists the key properties of a fundamental particle, including its electric charge, spin, mass and mean lifetime, see Figure \ref{fig:UpQuark}. These values are taken from the Particle Data Group \cite{ParticleDataGroup2024Review} and related sources \cite{Karshenboim2004Precision, Quadt2007Top}. 
Students compare these quantities and learn to identify which group each particle belongs to, distinguishing among quarks, leptons, and bosons. 
During play, students also encounter essential ideas such as the Pauli Exclusion Principle, anti-matter and annihilation. \\

The higher levels incorporate many other particle physics concepts. These concepts include the fundamental forces and exchange particles (level 2), color charge (level 3), randomness and probability (level 4), $W^+$ and $W^-$ transformations (level 5), neutrino oscillations (level 6), and spontaneous decay/transformation (level 7). Baryons, mesons and other particle systems such as positronium and hydrogen serve as targets that players attempt to build. \\

The game is designed for upper secondary students and takes about ten minutes per round. In a typical one hour lesson, students can complete the first two levels, replaying them to reinforce understanding before advancing to higher levels. \\

Particle Builder makes use of the anti-colour convention proposed by Wiener in 2017 \cite{Wiener2017Alternative}. Research has shown that younger students often hold misconceptions about particle physics, including imagining particles as static objects, or assuming that particles have the same properties as macroscopic matter such as colour, temperature, or even personification \cite{Wiener2015Can}. To minimize these misunderstandings, the Particle Builder illustrations intentionally avoid representing the fundamental particles as spheres. Instead, the cards feature abstract, imaginative characters to emphasize that the images are symbolic rather than realistic. \\

\begin{figure}[!htbp]
	\centering
	\includegraphics[width=0.5\textwidth]{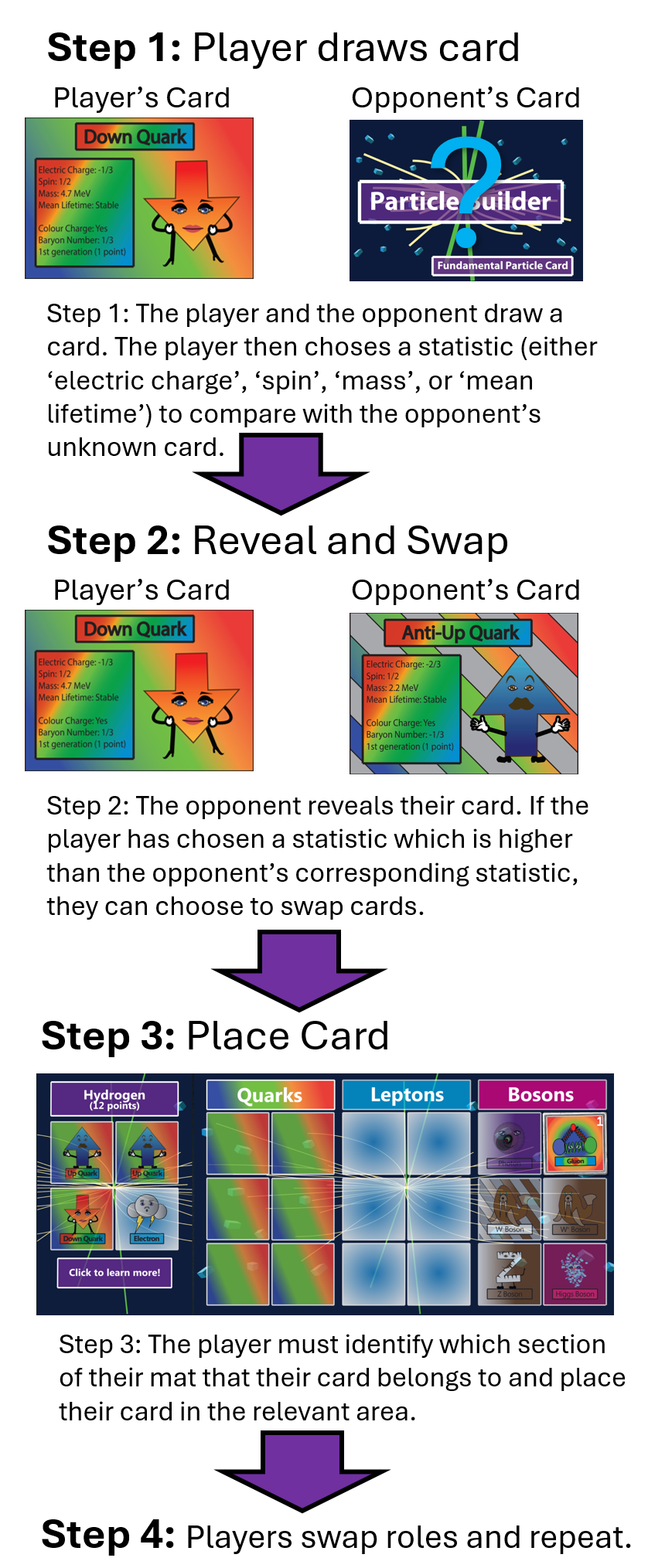}%
	\caption{\label{fig:Gameplay} Flow diagram showing the basic gameplay of Particle Builder. Full instructions for playing the game can be found at Zenodo \cite{McGinness2018Particle}.}
\end{figure}
\begin{figure*}[!t]
    \centering
    \includegraphics[width=0.85\textwidth]{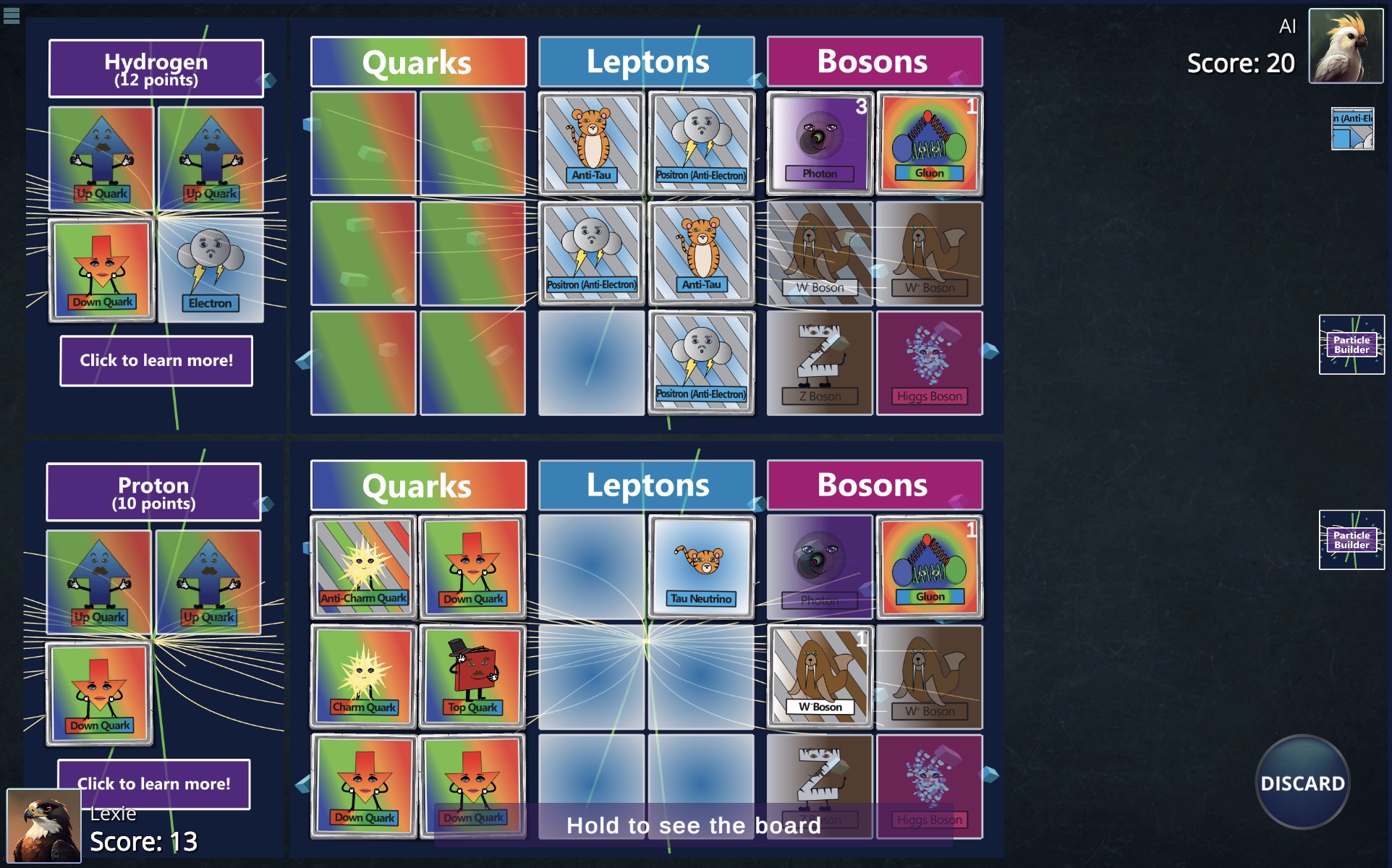}
    \caption{\label{fig:OnlineGame} Screenshot of the online version of Particle Builder, showing a human player versus the AI opponent.}
\end{figure*}

The online version was developed by students at the Australian National University and maintains the structure of the original board game. 
The online platform provides a tutorial to help students learn to play. 
Many studies have found that digital and gamified learning environments can increase student engagement and concentration while reducing learning-related stress \cite{Majuri2018Gamification, Lee2015Designing, Woo2014Digital, Halim2020Pupils, Cai2020SixStep}.
Figure \ref{fig:Gameplay} shows the basic gameplay flow, while Figure \ref{fig:OnlineGame} presents a screenshot of the interface. 

\section{Learning Outcomes}
Particle builder covers learning outcomes included in many high school curricula such as \cite{KranjcHorvat2022What,ACARA2024Physics,Norton2004Particle}: 
\begin{itemize}
	\setlength\itemsep{0.0em}
	\item recognize and name the six flavors of lepton and the six flavors of quarks 
	\item understand that all leptons and quarks have corresponding antiparticles 
	\item appreciate that quarks and anti-quarks combine to form baryons, anti-baryons and mesons 
	\item write balanced strong interactions, understanding the role of gluons 
	\item write balanced weak interactions, understanding the role of W and Z bosons 
\end{itemize}


Particle Builder directly supports these learning goals through its level-based design:
\begin{itemize}
	\setlength\itemsep{0.0em}
	\item Level 1: Students identify particle types (quarks, leptons, and bosons), and compare their electric charge, spin, mass, and lifetime. They also learn to recognize particle-antiparticle pairs and understand annihilation. 
	\item Level 2: Students explore the fundamental forces and their exchange particles (photon, gluon, Z boson). 
	\item Level 3: Students investigate color charge by assigning color to quarks and combining them into color neutral systems. 
	\item Level 4: Students are introduced to randomness through probabilistic interaction outcomes.
	\item Level 5: Students apply conservation of charge, baryon number, and lepton family number through $W^+$ and $W^-$ transformations.
	\item Level 6: Students simulate neutrino oscillations by allowing flavor changes among neutrinos.
	\item Level 7: Students investigate particle decays. 
\end{itemize}

\section{Survey Methods}

We wrote a set of nine questions which take students on average 5-6 minutes to complete as a pre and post survey. The survey covers the core topics covered in more than 60\% of curricula \cite{KranjcHorvat2022What}, focusing on \textit{anti-matter and annihilation}, \textit{leptons and quarks}, \textit{particle interactions}, and \textit{particle systems}. This was given to the students a few days prior to activity. The same questions were given again as a post-test afterwards. No other particle physics instruction was given between the surveys.\\



We had two control groups, where we evaluated the gain from 1.5 weeks (approximately 7 hours) of particle physics instruction. One control group ($N=23$) received traditional instruction (powerpoints and worksheets), while the other ($N=7$) received traditional instruction but completed the quark-puzzle activity \cite{McGinness2019ThreeD} in the place of one traditional lesson. We use this as a baseline measure but note that the improvement in the controls is for more than a week's worth of instruction, not just a single activity or lesson.\\


The game was taken to secondary school classes, STEM clubs and STEM camps in Australia, mostly in Canberra. Although there was a focus on Year 12 participants there were some participants as young as Year 7 in some of the groups. The researchers came to the lesson or event either with copies of the physical board game or the students used their own devices to play the online game. In total there were 281 students who completed the pre-test and 225 students who completed both the pre-test and the post-test.\\

Of the 225 students, 118 were male, 84 were female and 23 did not identify. 111 of the students were in Year 12 with the other 114 in younger secondary school year levels. Australia has three main schooling systems, the number of students from each was: $>53$ from public schools, $>24$ from systemic Catholic schools and $>51$ from private schools. School type was not recorded for the STEM clubs or extension classes, so we can put a lower limit of the number of students from each of the systems. In total 97 students played the physical game, 89 the online game and 42 played both.


\section{Results}

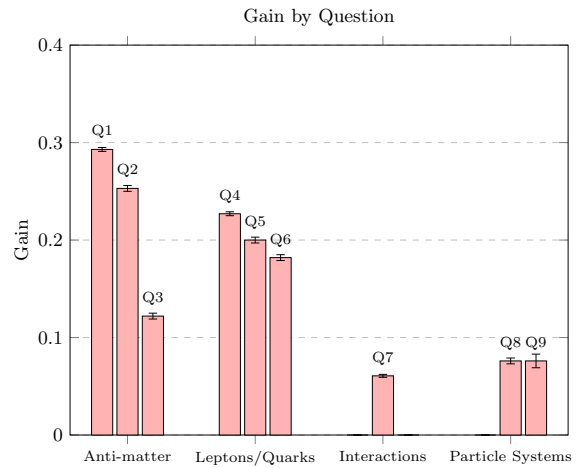
\begin{figure}
	\centering
    \scalebox{0.8}{
	\begin{tikzpicture}
		\begin{axis}[
			title={Gain by Question},
			ybar,
			bar width=10pt,
			width=0.55\textwidth,
			height=0.45\textwidth,
			ymin=0, ymax=8,
			xlabel={},
			ylabel style={rotate=0, yshift=-12pt},
			ylabel={Gain},
			ymin=0, ymax=0.4,
			ytick={0, 0.1, 0.2, 0.3, 0.4}, 
			yticklabels={0, 0.1, 0.2, 0.3, 0.4 }, 
			xticklabel style={font=\scriptsize}, 
			symbolic x coords={Anti-matter, Leptons/Quarks, Interactions, Particle Systems},
			xtick=data,
            nodes near coords,
            point meta=explicit symbolic,
            nodes near coords align={vertical},
            every node near coord/.append style={font=\scriptsize, xshift=0pt, yshift=1pt, anchor=south},
			legend style={at={(0.5,-0.15)},
				anchor=north,legend columns=-1},
			enlarge x limits=0.15,
			ymajorgrids,
			grid style=dashed,
			error bars/y dir=both,
			error bars/y explicit,
			every node near coord/.append style={font=\footnotesize}
			]
			
			\addplot+[
			ybar,
			color=black,
			fill=red!30,
			error bars/.cd,
			y dir=both,
			y explicit,
			] plot coordinates {
				(Anti-matter, 0.293) +- (0,0.002) 
				(Leptons/Quarks, 0.227) +- (0,0.002) 
				(Interactions, 0) +- (0,0) 
				(Particle Systems, 0) +- (0,0)
			};
			
			\addplot+[
			ybar,
			color=black,
			fill=red!30,
			error bars/.cd,
			y dir=both,
			y explicit,
			] plot coordinates {
				(Anti-matter, 0.253) +- (0,0.003) 
				(Leptons/Quarks, 0.200) +- (0,0.003) 
				(Interactions, 0.0607) +- (0,0.0015) 
				(Particle Systems, 0.076) +- (0,0.003)  
			};
			
			\addplot+[
			ybar,
			color=black,
			fill=red!30,
			error bars/.cd,
			y dir=both,
			y explicit,
			] plot coordinates {
				(Anti-matter, 0.122) +- (0,0.003) 
				(Leptons/Quarks, 0.182) +- (0,0.003) 
				(Interactions, 0) +- (0,0)
				(Particle Systems, 0.076) +- (0,0.007) 
			};

		\node[anchor=south, font=\scriptsize, xshift=-12, yshift=2pt] at (axis cs:Anti-matter,0.293) {Q1};
        \node[anchor=south, font=\scriptsize, xshift=-12, yshift=2pt] at (axis cs:Leptons/Quarks,0.227) {Q4};
        
        \node[anchor=south, font=\scriptsize, yshift=2pt] at (axis cs:Anti-matter,0.253) {Q2};
        \node[anchor=south, font=\scriptsize, yshift=2pt] at (axis cs:Leptons/Quarks,0.200) {Q5};
        \node[anchor=south, font=\scriptsize, yshift=2pt] at (axis cs:Interactions,0.0607) {Q7};
        \node[anchor=south, font=\scriptsize, yshift=2pt] at (axis cs:Particle Systems,0.076) {Q8};
        
        \node[anchor=south, font=\scriptsize, xshift=12, yshift=2pt] at (axis cs:Anti-matter,0.122) {Q3};
        \node[anchor=south, font=\scriptsize, xshift=12, yshift=2pt] at (axis cs:Leptons/Quarks,0.182) {Q6};
        \node[anchor=south, xshift=12, font=\scriptsize, yshift=2pt] at (axis cs:Particle Systems,0.076) {Q9};
		\end{axis}
	\end{tikzpicture}
    }
	\caption{Each column corresponds to the gain scores for a question from the pre/post-test. Errors bars indicate the standard error in the mean from the $n=225$ students.}
	\label{fig:QuestionBreakdown}
\end{figure}

In Physics Education Research the gain is a standard measure used to determine the level of student learning when pre/post-testing is used \cite{McGinness2014Development}. 
Gain can be interpreted as the increase in the fraction of questions which are answered correctly by students.  
The average gain across the entire 225 students was $0.16$. This is comparable to the gain from the traditional instruction group ($0.17$), but less than the group which completed the quark puzzle in their 1.5 weeks of instruction ($0.23$). This preliminary analysis shows that playing particle builder for a double lesson (1-2 hours) results in student learning comparable to a week and a half (approximately 7 hours) of traditional teaching. 
Figure \ref{fig:QuestionBreakdown} shows the gain for each individual question, divided by sub-topic. The greatest improvement for students were on the \textit{anti-matter} and \textit{leptons and quarks} topics which correspond directly to game mechanics in Level 1. \\

In our post-activity survey, we follow the convention used in previous studies which asks students to self-report their learning, enjoyment, and engagement/motivation \cite{McGinness2014Development, McGinness2016Action}. For this, we ask the students to compare the Particle Builder activity to a normal classroom lesson. Overall, 94\% of students found the game more enjoyable than a normal science lesson and 88\% found it more engaging. 68\% of students reported that they learned more in this activity than in a normal lesson and only 32\% of students found it more difficult. Feedback from students clearly indicates that they enjoyed the game, but preferred the physical board game to the online game. 
However it is significantly less effort for teachers to use the online game as they do not need to print and cut out the cards and game mats.

\section{Conclusion}

Particle Builder was shown to have significant educational benefit for students, leading to consistent improvement on basic particle physics questions. Students almost universally reported that the activity was more engaging, more enjoyable, less difficult and better for their learning than normal classroom activities. 
Both the physical and online versions of Particle Builder are freely available for instructors to use in their classrooms \cite{McGinness2018Particle, McGinness2024Particle}.

\section{Acknowledgements}
\begin{acknowledgments}

We would like to thank Harri Leinonen who worked with Lachlan McGinness to create the first version of Particle Builder. We also thank and acknowledge Rowan McGinness who developed all of the art used in the game.\\

The ethical aspects of this research have been approved by the ANU Human Research Ethics Committee (Protocol 2024/0566).

\end{acknowledgments}



\end{document}